\documentclass{PoS}

\usepackage{amsmath}
\usepackage{amssymb}
\usepackage{amsthm}
\usepackage{graphicx}
\usepackage{url}

\newcommand{\mi}{\mathrm{i}} 

\newcommand{\dfd}[3]{\hspace{-0.4em}\ensuremath{\frac{\mathrm{d}^{#1}#3}{(2\pi)^{#2}}}\,}

\newcommand{\eqn}[1]{Eq.~(\ref{#1})}
\newcommand{\fig}[1]{Figure~\ref{#1}}
\newcommand{\tab}[1]{Table~\ref{#1}}

\def\Kbar{\overline{K}}
\def\K0bar{\overline{K^0}}

\title{
  \hfill {\small LFTC-17-20/20} \\
  $\phi$ meson in nuclear matter and nuclei}

\ShortTitle{$\phi$-meson in nuclear matter and nuclei}

\author{\speaker{J.~J.~Cobos-Mart\'{\i}nez}$^{ab}$, K.~Tsushima$^{b}$,
  G.~Krein$^{c}$, and A.~ W.~Thomas$^{d}$ \\
\llap{$^{a}$} CONACyT--Departamento de F\'{\i}sica, Centro de
Investigaci\'on y de Estudios Avanzados del Instituto Polit\'ecnico
Nacional, Apartado Postal 14-740, 07000, Ciudad de M\'exico, M\'exico. \\
\llap{$^{b}$} Laborat\'orio de F\'{\i}sica Te\'orica e Computacional-LFTC,
Universidade Cruzeiro do Sul, 01506-000, S\~ao Paulo, SP, Brazil. \\
\llap{$^{c}$} Instituto de F\'{\i}sica Te\'orica, Universidade Estadual
Paulista, Rua Dr. Bento Teobaldo Ferraz, 271-Bloco II, 01140-070, S\~ao
Paulo, SP, Brazil. \\
\llap{$^{d}$} ARC Centre of Excellence for Particle Physics at the Terascale
and CSSM, Department of Physics, University of Adelaide, Adelaide SA 5005,
Australia. \\
  
E-mail: \email{jcobos@fis.cinvestav.mx}, \email{kazuo.tsushima@gmail.com},
\email{gkrein@ift.unesp.br}, \email{anthony.thomas@adelaide.edu.au}}

\abstract{The mass and decay width of the $\phi$ meson in cold nuclear
  matter are computed in an effective Lagrangian approach, with the medium
  dependence of these properties obtained by evaluating kaon-antikaon loop
  contributions to the $\phi$ self-energy.
  At normal nuclear matter density, we find a downward shift of the $\phi$
  mass by a few percent, while the decay width is enhanced by an order of
  magnitude. We also present $\phi$--nucleus bound state energies and
  absorption widths for some selected nuclei, using complex optical potentials
  obtained in the local density approximation.
  Our results suggest that the $\phi$ should form bound states with all the
  nuclei considered. However, the identification of the signal for these
  predicted bound states will need careful investigation because of their
  sizable absorption widths.}

\FullConference{XVII International Conference on Hadron Spectroscopy and
  Structure \\
  25-29 September, 2017 \\
  University of Salamanca, Salamanca, Spain}

\begin{document}

\section{Introduction}
\vspace{-1.25mm}

The properties of the $\phi$ meson at finite baryon density, such as its mass
and decay width, have attracted considerable experimental and theoretical
interest over the last few decades~\cite{vectormesonsinnuclmatt}.
This has been in part due to its potential to carry information on the partial
restoration of chiral symmetry, the possible role of QCD of van der Waals
forces in the binding of quarkonia to
nuclei~\cite{vanderwaals}, and the strangeness content of the
nucleon~\cite{strangenesscontent}.
However, a experimental unified consensus has not yet been reached among the
different experiments concerning the $\phi$. A large in-medium
broadening of its decay width has been reported by most of the experiments
performed, while only a few of them find evidence for a substantial mass
shift~\cite{phipptiesnuclmatt}.
The study of the $\phi$--nucleus bound states~\cite{JPARCE29Proposal,
  JLabphiProposal} is complementary to the invariant mass measurements, where
only a small fraction of the produced $\phi$ decay inside the nucleus, and
may be expected to provide extra information on the $\phi$ properties at
finite baryon density, since a downward mass shift of the $\phi$ in a nucleus
is directly connected with the existence of an attractive potential between
the $\phi$ and the nucleus where it has been produced.
This new experimental approach will produce a slowly moving $\phi$ such that
the maximum nuclear matter effect can be probed. 
On the theoretical side, various authors predict a small downward shift of the
in-medium $\phi$ mass and a large broadening of its decay
width~\cite{phipptiestheory} at normal nuclear matter density.
An important question is whether this attraction, if it exists, is sufficient
to bind the $\phi$ to a nucleus.
Using simple quantum mechanical arguments, the prospects of capturing a
$\phi$ seem quite favorable, provided that the $\phi$ can be produced almost
at rest in the nucleus. However, a full calculation using realistic density
profiles of nuclei is required for a more reliable estimate.
The work presented at this conference has been published in
Refs.~\cite{Cobos-Martinez:2017vtr,Cobos-Martinez:2017woo}. In Ref.~\cite{Cobos-Martinez:2017vtr} we computed the $\phi$ mass shift and decay width in
nuclear matter by evaluating the $K\overline{K}$ loop contribution to the
$\phi$ self-energy, with the in-medium $K$ and $\overline{K}$ masses
explicitly calculated using the quark-meson coupling (QMC)
model~\cite{Tsushima:1997df}. This initial study is been extended
in Ref.~\cite{Cobos-Martinez:2017woo} to some selected nuclei by computing the
$\phi$--nucleus bound state energies and absorption with a complex optical
potential calculated in the local density approximation.

\section{$\phi$ mass and decay width in nuclear matter}

\begin{figure}
\begin{center}
\scalebox{0.63}{
\begin{tabular}{ccc}
 \includegraphics[scale=0.2]{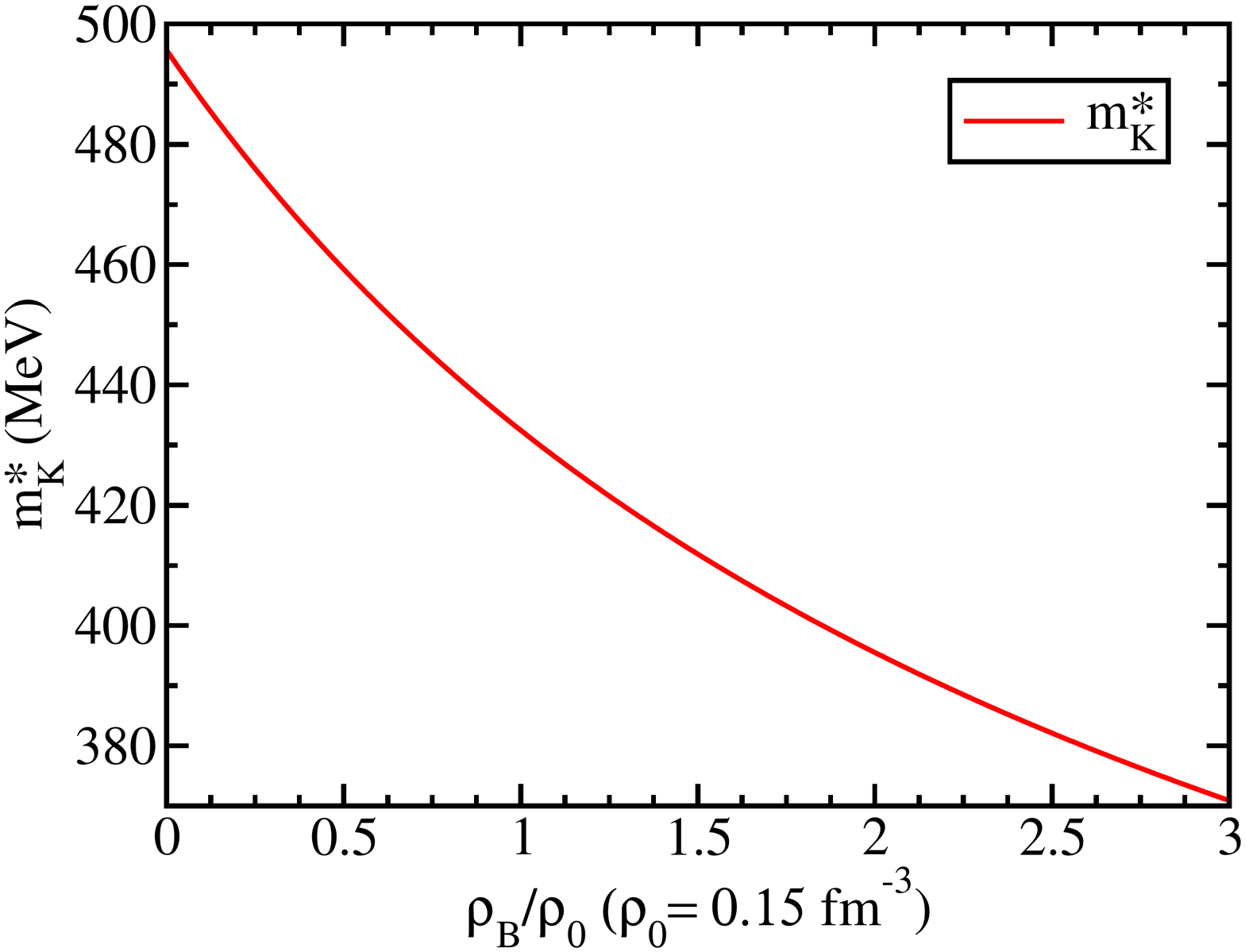} &
 \includegraphics[scale=0.2]{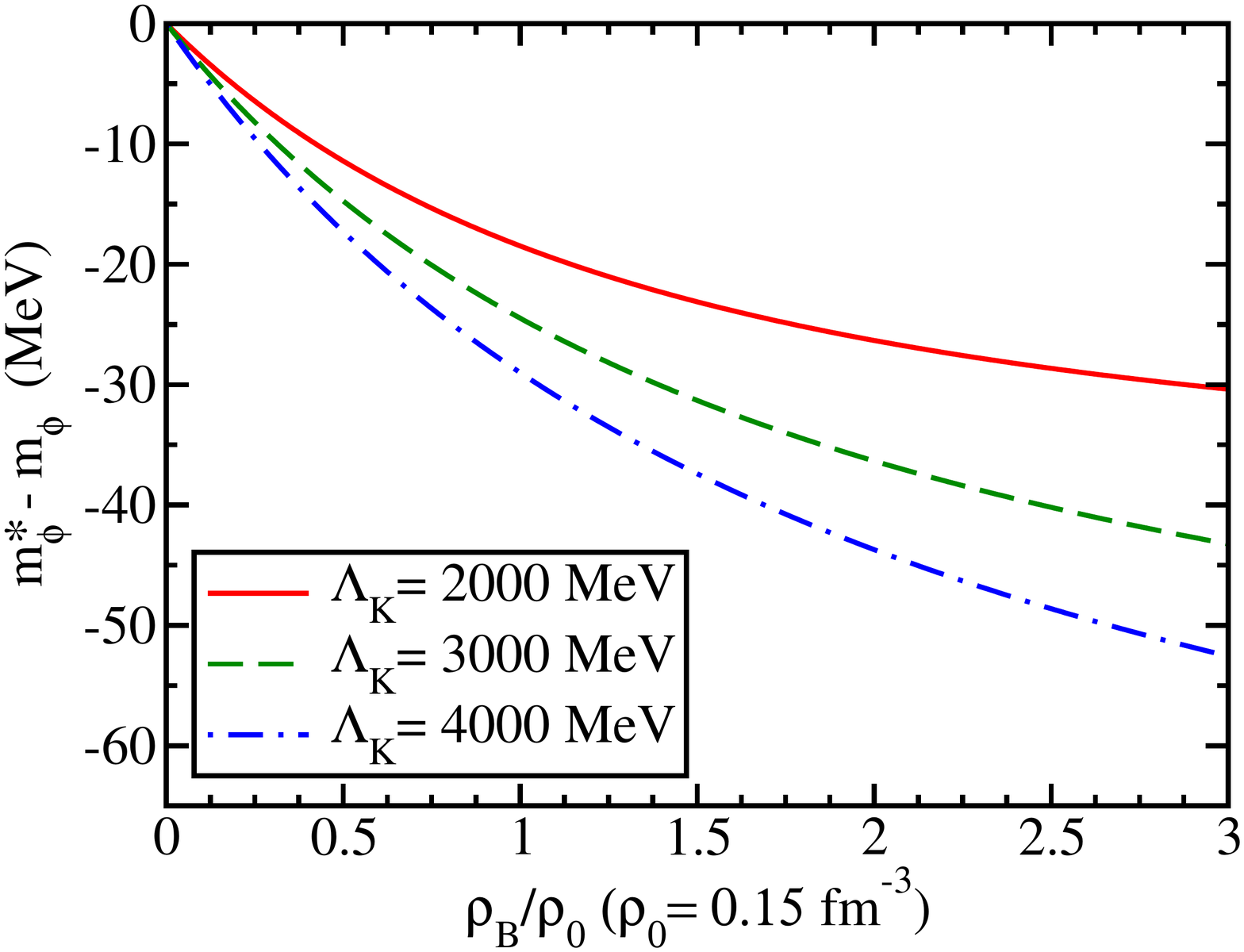} &
 \includegraphics[scale=0.176]{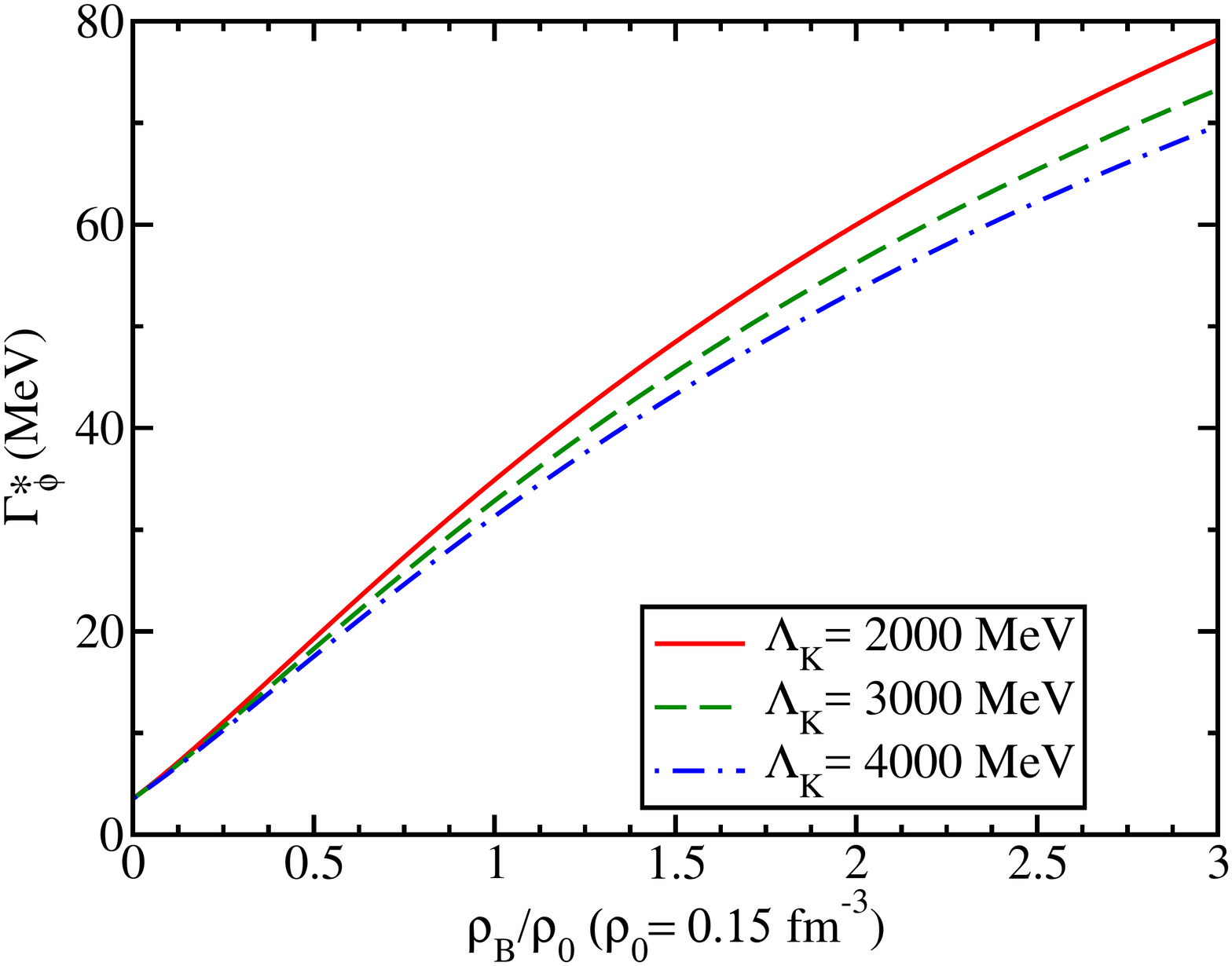} \\
\end{tabular}
}
\caption{\label{fig:nuclmatt} Left panel: In-medium Lorentz scalar
  kaon (=antikaon) mass $m_{K}^{*}$; centre and right panels: $\phi$ mass
  shift and decay width, respectively, for three
  values of $\Lambda_{K}$.}
\end{center}
\end{figure}

One expects that a significant fraction of the density dependence of
the $\phi$ self-energy in nuclear matter arises from the in-medium
modification of the $K\Kbar$ intermediate state in the $\phi$ self-energy,
since this is the dominant decay channel in vacuum.
We briefly review the computation~\cite{Cobos-Martinez:2017vtr} of the
$\phi$ self-energy in vacuum and in nuclear matter using an effective
Lagrangian approach~\cite{Klingl:1996by}.
The interaction Lagrangian contains only the $\phi K\overline{K}$ vertex, and
is given by
\begin{equation}
\label{eqn:Lpkk}
\mathcal{L}_{\phi K\overline{K}}=\mi g_{\phi}\phi^{\mu}
\left[\Kbar(\partial_{\mu}K)-(\partial_{\mu}\Kbar)K\right],
\end{equation}
\noindent where $K$ and $\Kbar$ are isospin doublets.
The contribution from the $\phi K\overline{K}$ coupling to the scalar part of
the $\phi$ self-energy, $\Pi_{\phi}(p)$, is given by 
\begin{equation}
\label{eqn:phise}
\mi\Pi_{\phi}(p)=-\frac{8}{3}g_{\phi}^{2}\int\dfd{4}{4}{q}\vec{q}^{\,2}
D_{K}(q)D_{K}(q-p),
\end{equation}
\noindent where $D_{K}(q)=1/\left(q^{2}-m_{K}^{2}+\mi\epsilon\right)$ is the
kaon propagator;  $p=(p^{0}=m_{\phi},\vec{0})$ for $\phi$ at rest, with
$m_{\phi}$ the $\phi$ mass; $m_{K} (=m_{\Kbar})$ is
the kaon mass, and $g_{\phi}= 4.539$~\cite{Cobos-Martinez:2017vtr} is the
coupling constant. The integral in \eqn{eqn:phise} is divergent but it will
be regulated using a phenomenological form factor, with cutoff parameter
$\Lambda_{K}$~\cite{Cobos-Martinez:2017vtr}.
The sensitivity of the results to the value of $\Lambda_{K}$ is analyzed below.
The mass and decay width of the $\phi$ in vacuum ($m_{\phi}$ and
$\Gamma_{\phi}$), as well as in nuclear matter ($m_{\phi}^{*}$ and
$\Gamma_{\phi}^{*}$), are determined~\cite{Cobos-Martinez:2017vtr} from
\begin{eqnarray}
\label{eqn:phimass} 
m_{\phi}^{2}&=&\left(m_{\phi}^{0}\right)^{2}+\Re\Pi_{\phi}(m_{\phi}^{2}), \\
\label{eqn:phiwidth}
\Gamma_{\phi}&=&-\frac{1}{m_{\phi}}\Im\Pi_{\phi}(m_{\phi}^{2}).
\end{eqnarray}
The density dependence of the $\phi$ mass and decay width is driven by the
intermediate $K\Kbar$ state interactions with the nuclear medium.
This effect enters through the Lorentz scalar mass $m_{K}^{*}$ in the kaon
propagators in \eqn{eqn:phise}.
The in-medium properties of $K$ and $\Kbar$ are calculated in the QMC
model~\cite{Tsushima:1997df} (see Ref.~\cite{Saito:2005rv} for a review)
considering infinitely large, uniformly symmetric, spin-isospin-saturated
nuclear matter in its rest frame, where all the scalar and vector mean field
potentials, which are responsible for the nuclear many-body interactions,
become constant in the Hartree approximation.
In \fig{fig:nuclmatt} (left panel) we present the resulting in-medium kaon
Lorentz scalar mass (=antikaon Lorentz scalar mass) as a function of the
baryon density. At normal nuclear matter density $\rho_{0}= 0.15$ fm$^{-3}$
the effective kaon mass  has decreased by about 13\%.
We recall that the isoscalar-vector $\omega$ mean field potentials arise both
for the kaon and antikaon, but they can be eliminated by a variable shift in
the loop calculation of the $\phi$ self-energy, and therefore are not show
them here.
In \fig{fig:nuclmatt}, we present the $\phi$ mass shift (centre panel)
and decay width (right panel) as a function of the nuclear matter density,
$\rho_{B}$, for three values of $\Lambda_{K}$. The effect of the in-medium
kaon and antikaon mass change yields a negative mass shift for the $\phi$.
This is because the reduction in the kaon and antikaon masses enhances the
$K\Kbar$ loop contribution in nuclear matter relative to that in vacuum.
For the largest value of $\rho_{B}$, the downward mass shift turns out to be
a few percent at most for all values of $\Lambda_{K}$. On the other hand,
$\Gamma_{\phi}^{*}$ is very sensitive to the change in the kaon and antikaon
masses, increasing rapidly with increasing nuclear matter density, up to a
factor of $\approx 20$ for the largest value of $\rho_{B}$.
These results open the experimental possibility for studying the binding and
absorption of $\phi$ in nuclei. Although the mass shift found in this study
may be large enough to bind the $\phi$ to a nucleus, the broadening of its
decay width will make it difficult to observe a signal for the $\phi$--nucleus
bound state formation experimentally. We explore this further in the second
part of this talk.

\section{$\phi$--nucleus bound states }

We now explore the situation where the $\phi$ meson is placed inside a
nucleus~\cite{Cobos-Martinez:2017woo}.
The nuclear density distributions for all nuclei but $^{4}$He are obtained
in the QMC model~\cite{Saito:1996sf}.
For $^{4}$He, we use Ref.~\cite{Saito:1997ae}.
Then, using a local density approximation the $\phi$--nucleus potentials for a nucleus $A$ is given by
\begin{equation}
\label{eqn:Vcomplex}
V_{\phi A}(r)= U_{\phi}(r)-\frac{\mi}{2}W_{\phi}(r),
\end{equation}
\noindent where $r$ is the distance from the center of the nucleus and 
$U_{\phi}(r)=m^{*}_{\phi}(\rho_{B}(r))-m_\phi$ and
$W_{\phi}(r)=\Gamma_{\phi}(\rho_{B}(r))$ are, respectively, the $\phi$ mass shift
and decay width inside nucleus $A$, with  $\rho_{B}(r)$ the baryon density
distribution of the given nucleus.
\begin{figure}
  \centering
\scalebox{0.63}{
\begin{tabular}{cccc}
  \includegraphics[scale=0.2]{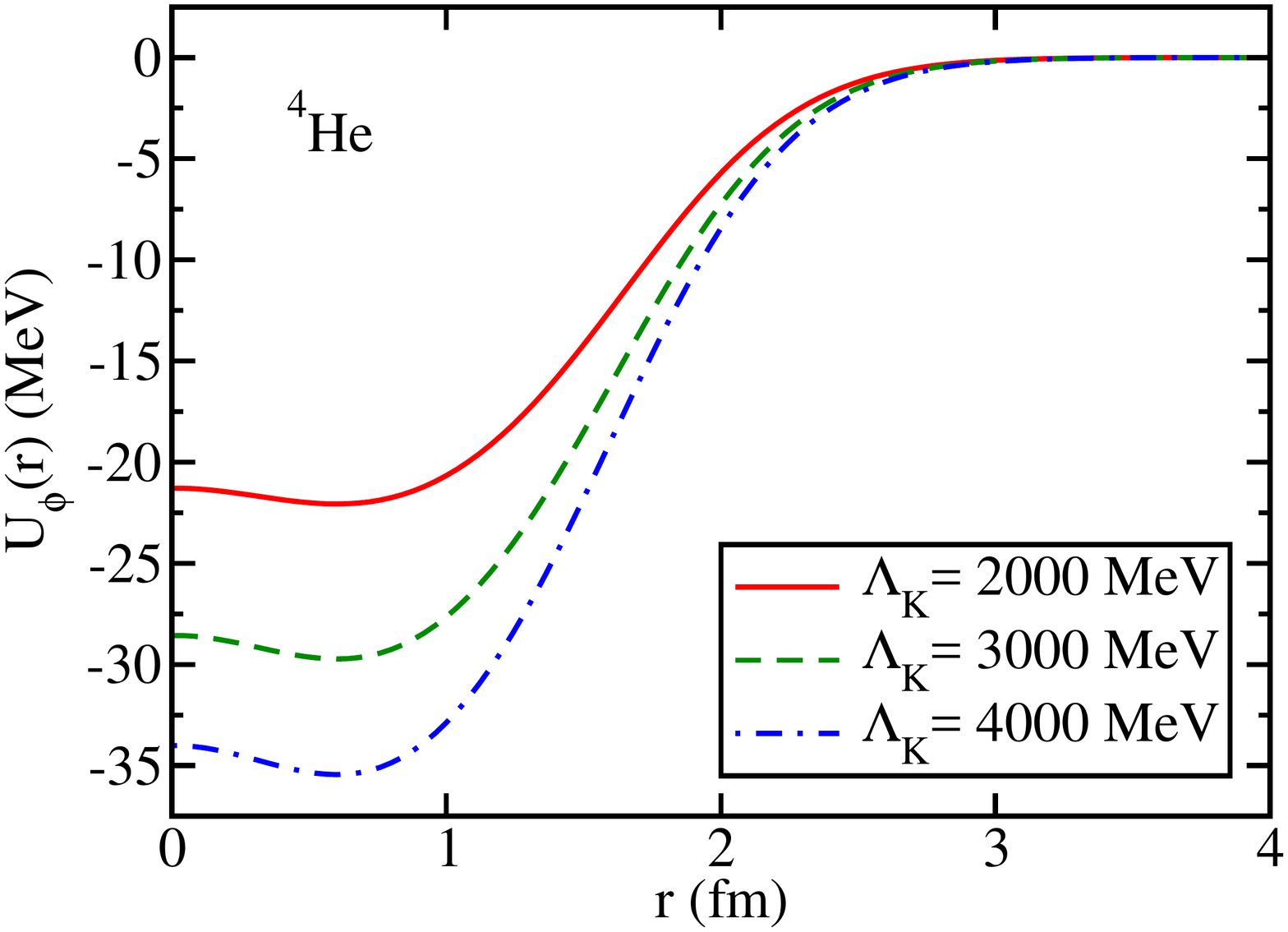} &
  \includegraphics[scale=0.2]{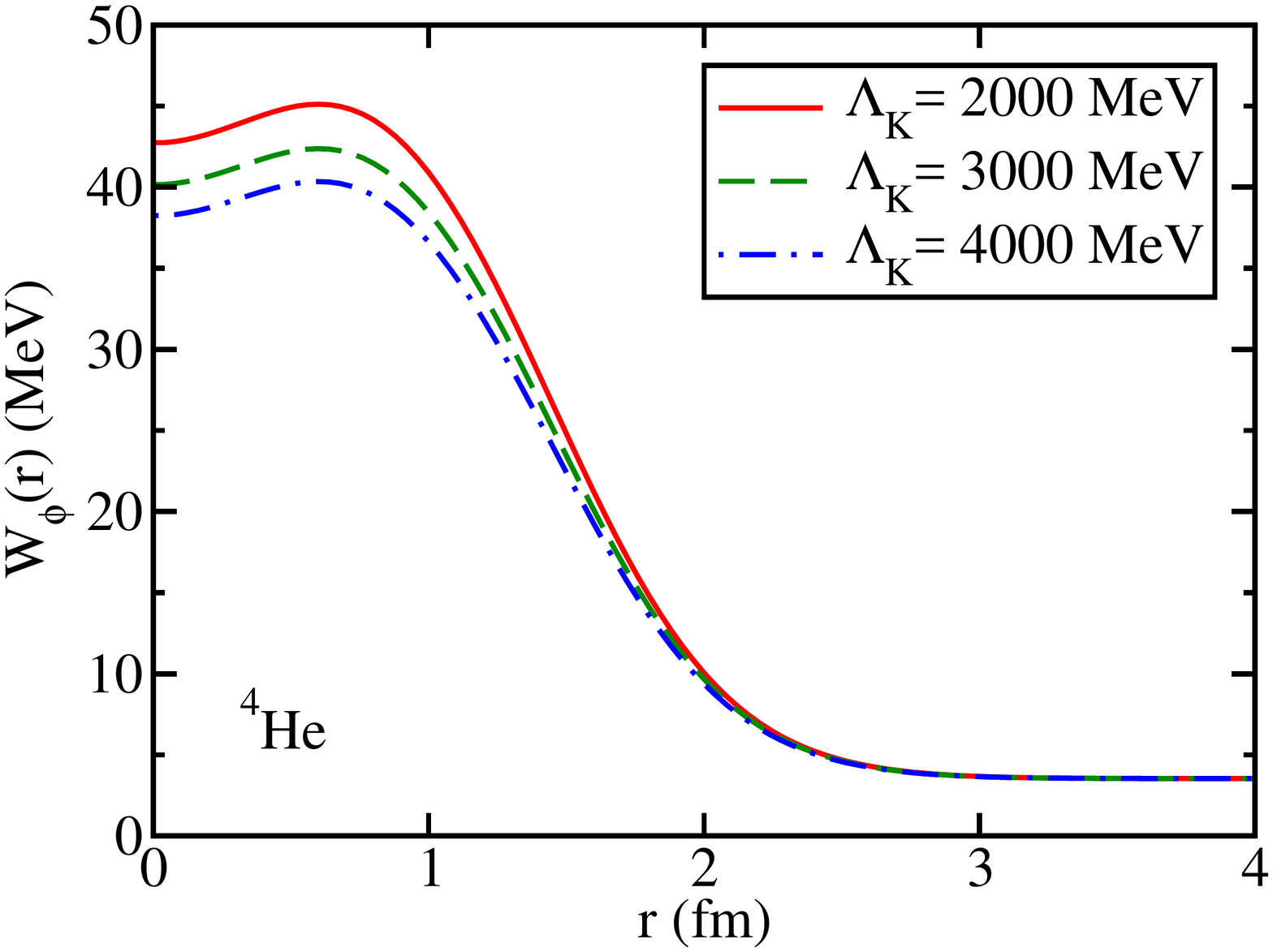} &
  \includegraphics[scale=0.2]{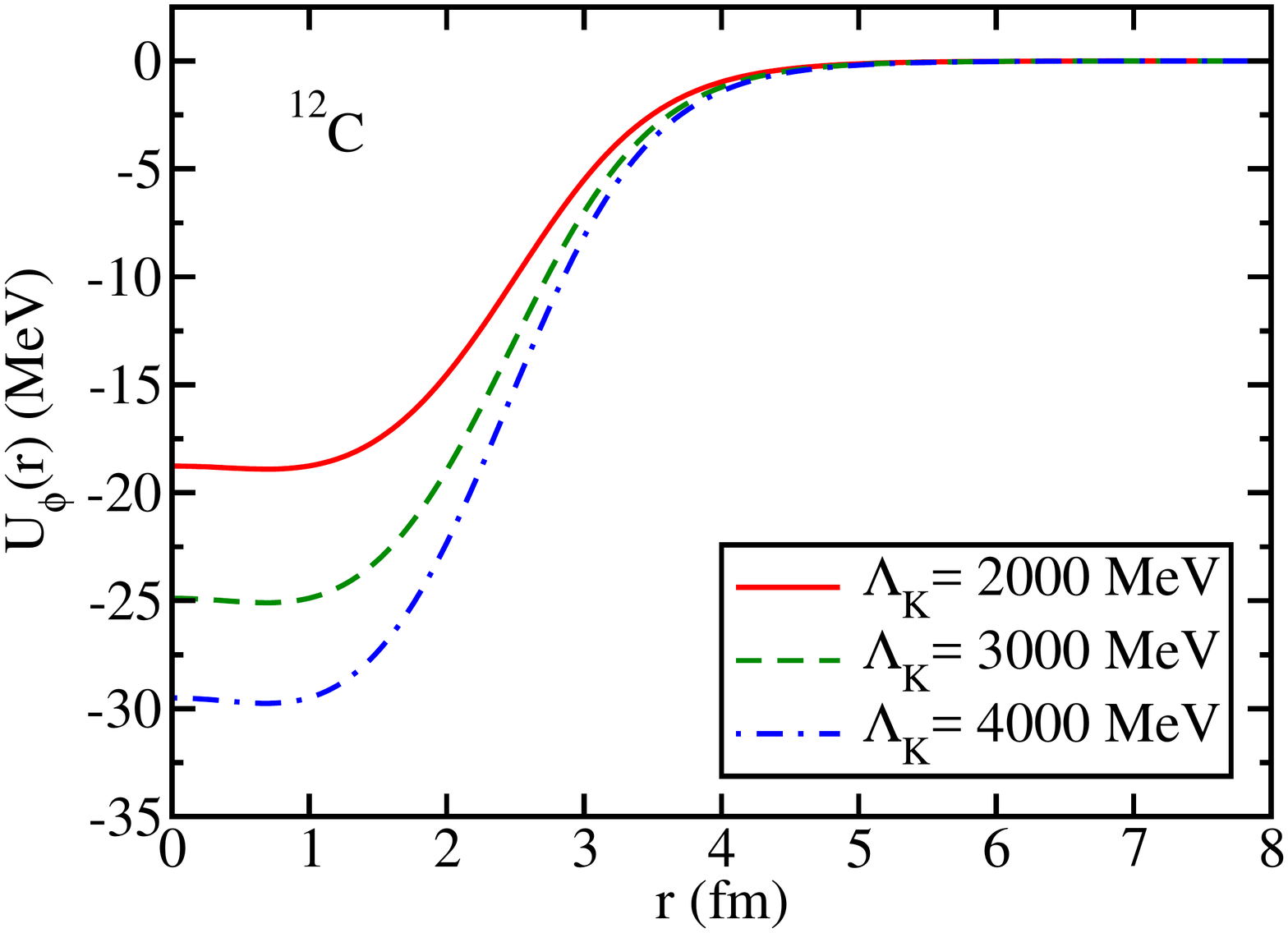} &
  \includegraphics[scale=0.2]{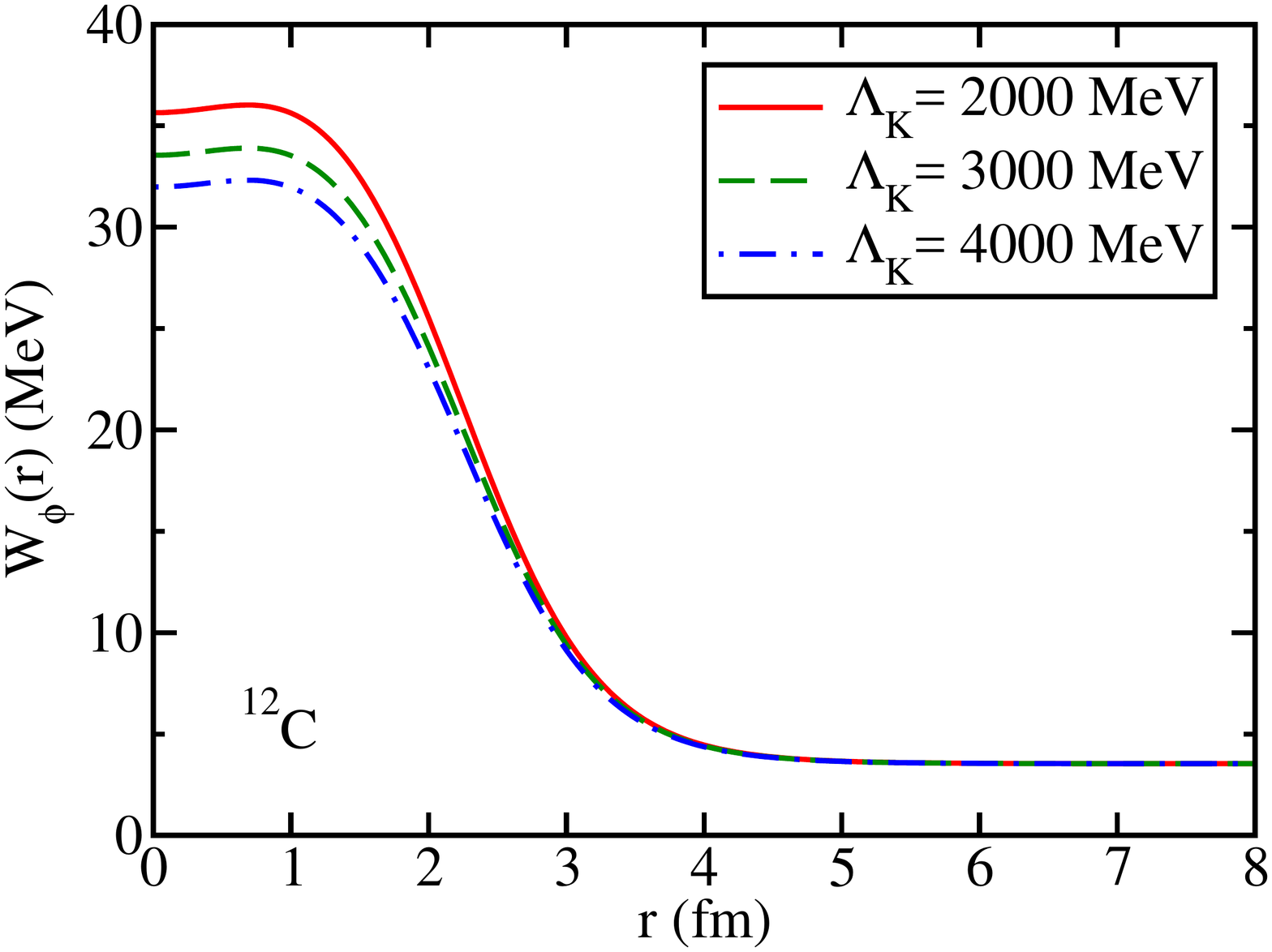} \\
  \includegraphics[scale=0.2]{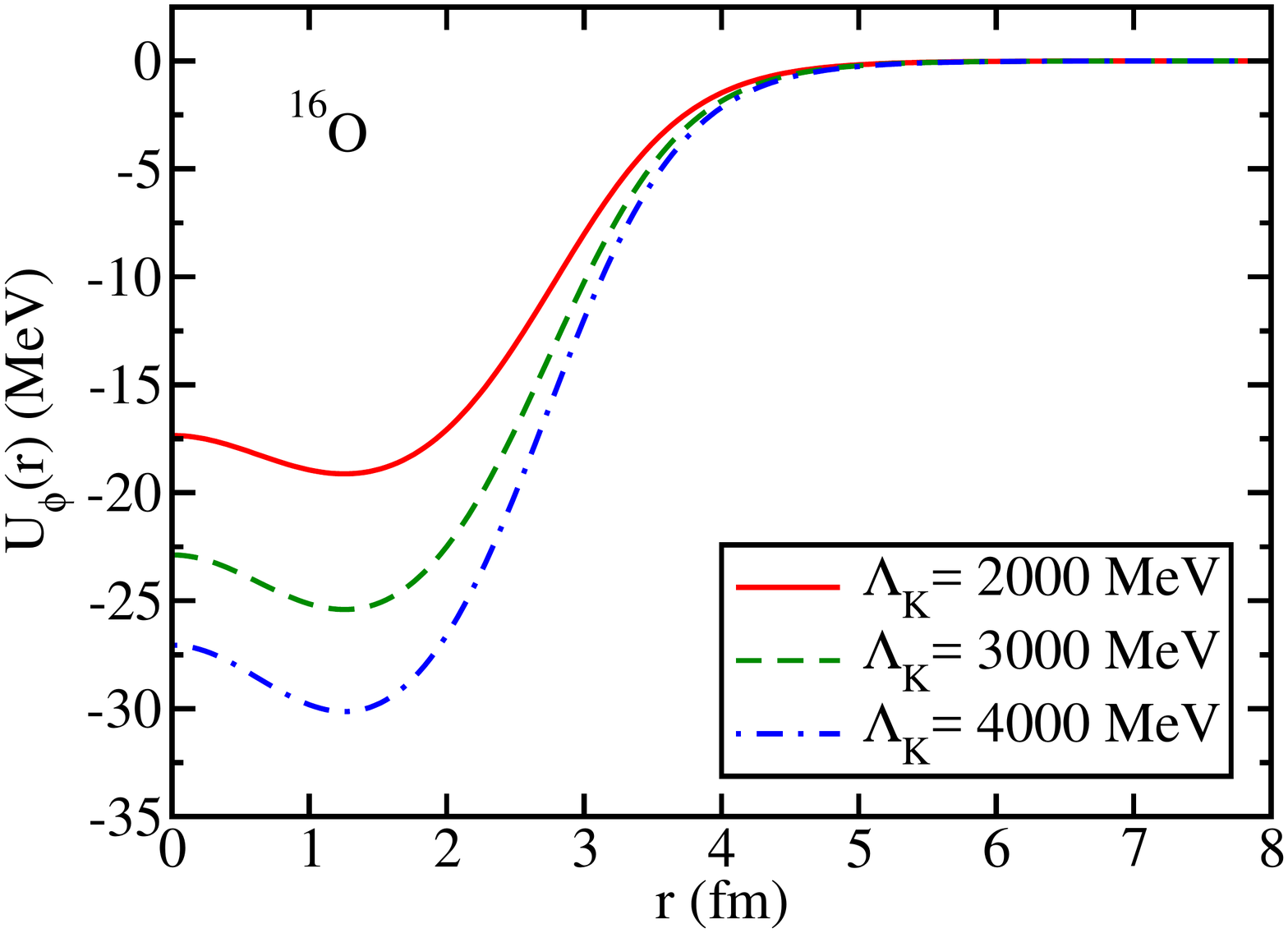} &
  \includegraphics[scale=0.2]{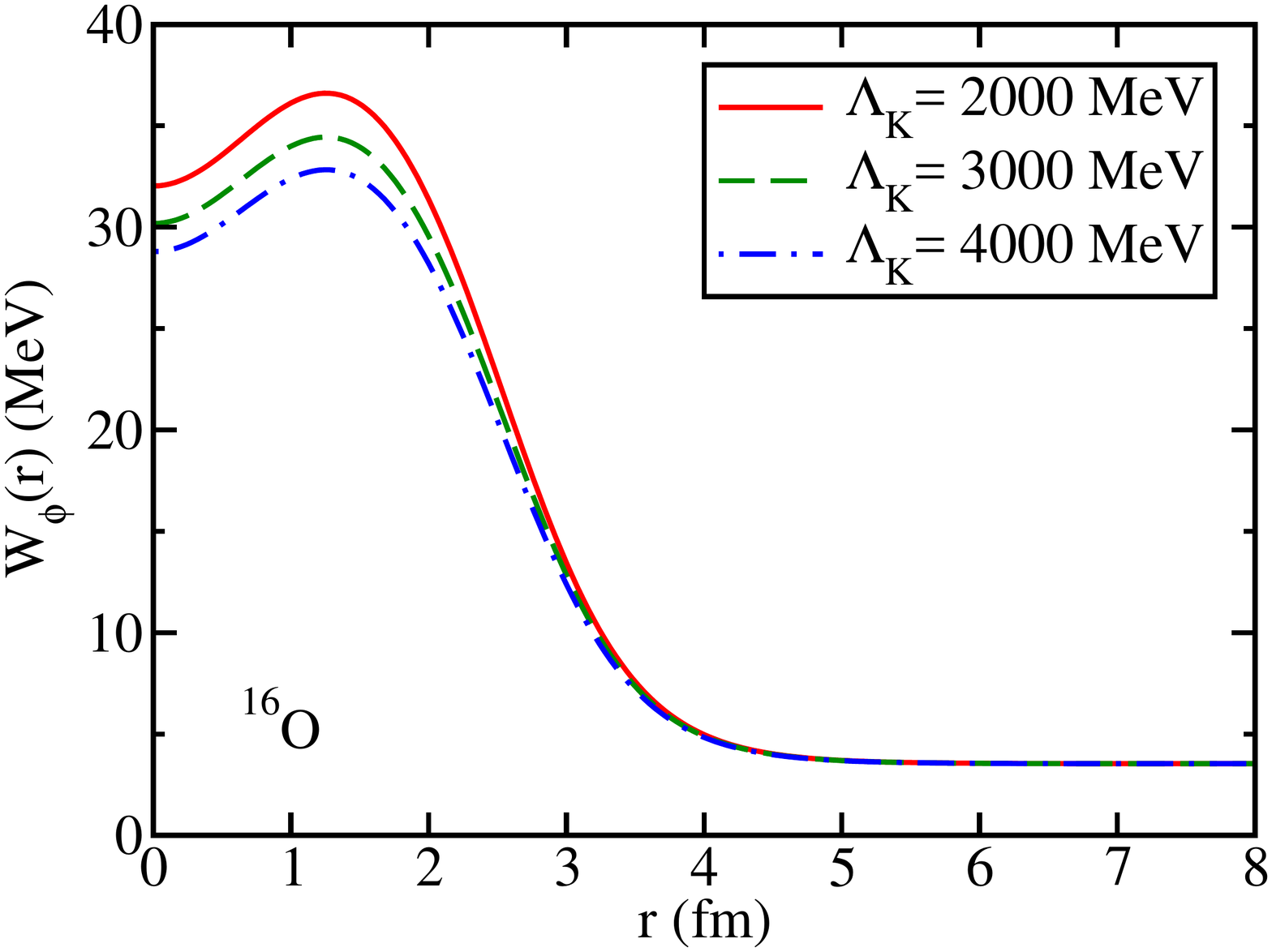} &
  \includegraphics[scale=0.2]{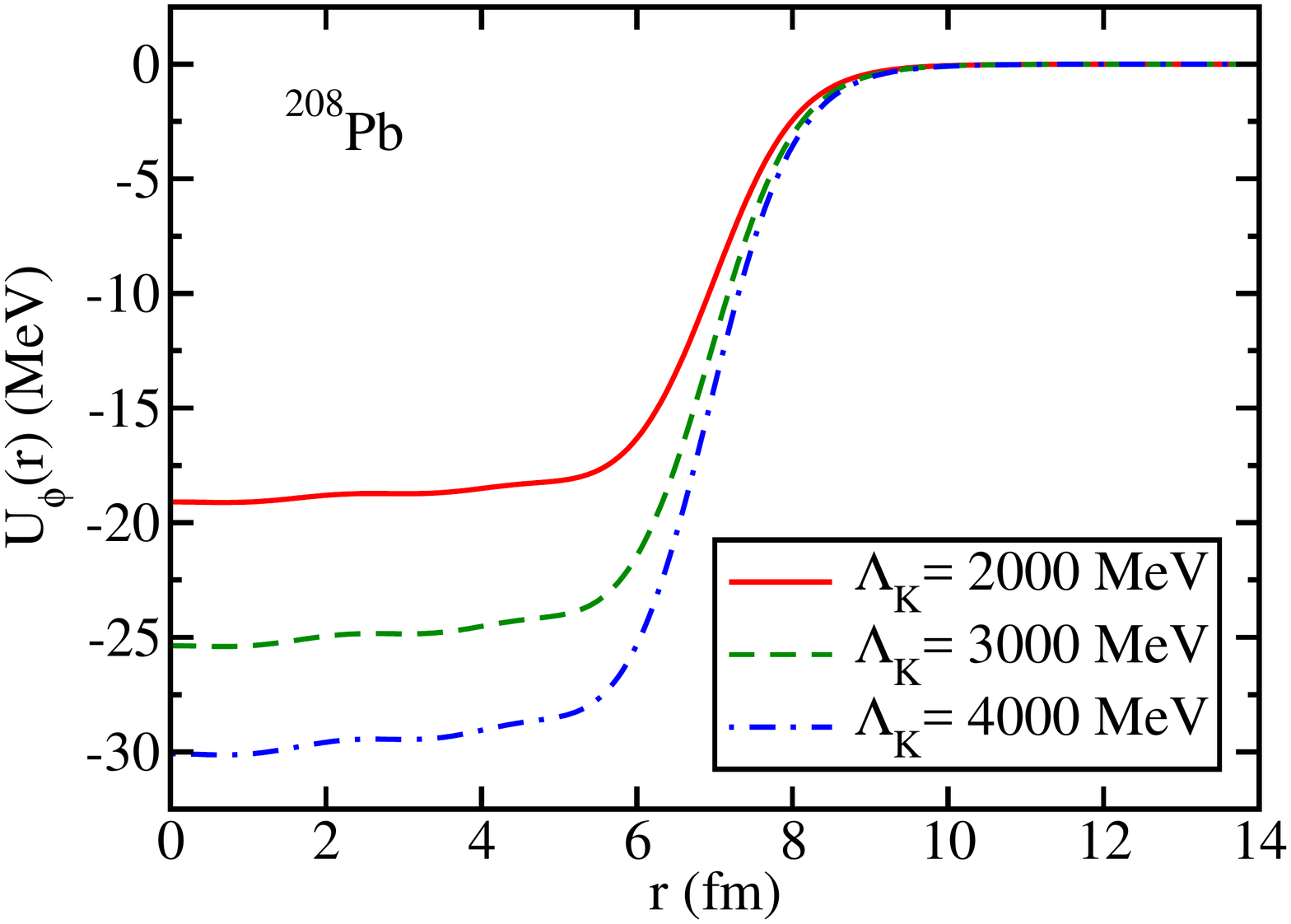} &
  \includegraphics[scale=0.2]{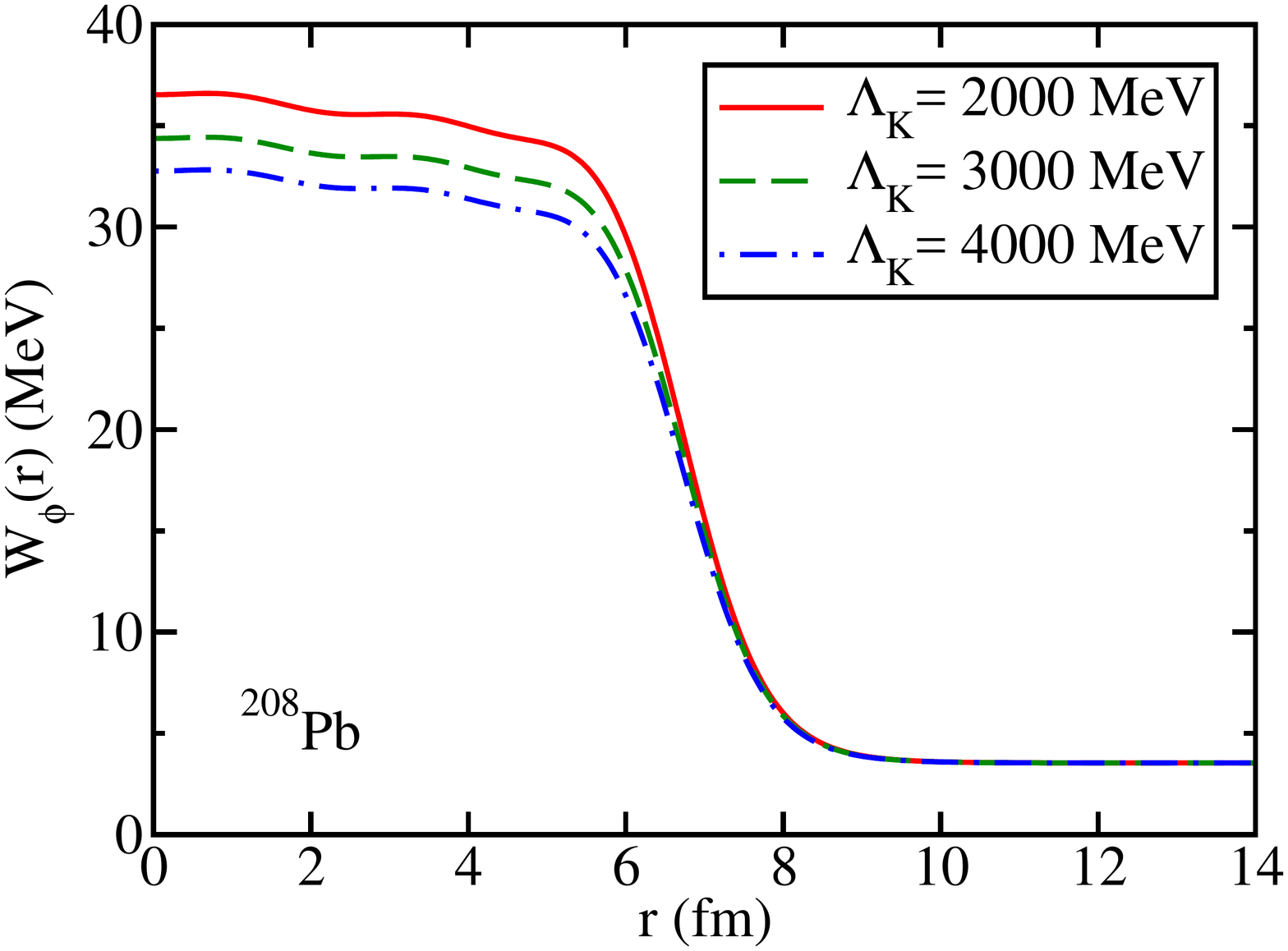} \\
\end{tabular}
}
\caption{\label{fig:phinuclpot} Real [$U_{\phi}(r)$] and imaginary [$W\phi(r)$]
  parts of the $\phi$--nucleus potentials for the nuclei studied here,
  for three values of $\Lambda_{K}$.}
\end{figure}
In \fig{fig:phinuclpot} we show the $\phi$ potentials
calculated~\cite{Cobos-Martinez:2017woo} for the selected nuclei. One can see
that the depth of the real part of the potential is sensitive to $\Lambda_{K}$.
On the other hand, the imaginary part does not vary much with $\Lambda_{K}$.
Using the $\phi$ potentials, we next calculate the $\phi$ single-particle
energies and absorption widths.
We consider the situation where the $\phi$ is produced nearly at
rest~\cite{Cobos-Martinez:2017woo}.
Then, under this condition, solving the Proca equation becomes equivalent to
solving the Klein-Gordon equation
\begin{equation}
\label{eqn:kge}
\left(-\nabla^{2} + \mu^{2} + 2\mu V(r)\right)\phi(\vec{r})
= \mathcal{E}^{2}\phi(\vec{r}),
\end{equation}
where $\mu$ is the reduced mass of the $\phi$--nucleus system in vacuum,
and $V (r)$ is given by \eqn{eqn:Vcomplex}. 
The calculated bound state energies ($E$) and absorption widths
($\Gamma$)~\cite{Cobos-Martinez:2017woo},
which are related to the complex energy eigenvalue $\mathcal{E}$ by
$E= \Re\mathcal{E}-\mu$ and $\Gamma= -2\Im\mathcal{E}$, are listed in
\tab{tab:phienergies} with and without the imaginary part of the potential.
When the imaginary part of the $\phi$--nucleus potential is set to zero
we see that the $\phi$ is expected to form bound states with all the nuclei
selected. The bound state energy is obviously dependent on $\Lambda_{K}$,
increasing as $\Lambda_{K}$ increases.
Adding the absorptive part of the potential changes the situation
considerably. We note that for the largest value of $\Lambda_{K}$, which
yields the deepest potentials, the $\phi$ is expected to form bound states
with all the selected nuclei, including $^{4}$He. However, in this case,
whether or not the bound states can be observed experimentally is sensitive
to the value of $\Lambda_{K}$.
One also observes that the width of the bound state is insensitive to the
values of $\Lambda_{K}$ for all nuclei.
Furthermore, since the so-called dispersive effect of the absorptive potential
is repulsive, the bound states disappear completely in some cases, even though
they were found when the absorptive part was set to zero. This feature is
obvious for the $^{4}$He nucleus, making it especially relevant to the future
experiments, planned at J-PARC and JLab using light and medium-heavy
nuclei~\cite{JPARCE29Proposal,JLabphiProposal}.
%
\begin{table}[ht]
\begin{center}
\scalebox{0.85}{
  \begin{tabular}{ll|rr|rr|rr} 
\hline \hline
& & \multicolumn{2}{c|}{$\Lambda_{K}=2000$} &
\multicolumn{2}{c}{$\Lambda_{K}=3000$} & 
\multicolumn{2}{|c}{$\Lambda_{K}=4000$}  \\
\hline
 & & $E$ & $\Gamma/2$ & $E$ & $\Gamma/2$ & $E$ & $\Gamma/2$ \\
\hline
$^{4}_{\phi}\text{He}$ & 1s & n (-0.8) & n & n (-1.4) & n & -1.0 (-3.2) & 8.3 \\
\hline
$^{12}_{\phi}\text{C}$ & 1s & -2.1 (-4.2) & 10.6 & -6.4 (-7.7) & 11.1 & -9.8
(-10.7) & 11.2 \\
\hline
$^{16}_{\phi}\text{O}$ & 1s & -4.0 (-5.9) & 12.3 & -8.9 (-10.0) & 12.5 & -12.6
(-13.4) & 12.4 \\
& 1p & n (n) & n & n (n) & n & n (-1.5) & n \\
\hline
$^{208}_{\phi}\text{Pb}$ & 1s & -15.0 (-15.5) & 17.4 & -21.1 (-21.4) & 16.6 &
-25.8 (-26.0) & 16.0 \\
& 1p & -11.4 (-12.1) & 16.7 & -17.4 (-17.8) & 16.0 & -21.9  (-22.2) & 15.5 \\
& 1d & -6.9 (-8.1) & 15.7 & -12.7 (-13.4) & 15.2 & -17.1 (-17.6) & 14.8 \\
& 2s & -5.2 (-6.6) & 15.1 & -10.9 (-11.7) & 14.8 & -15.2 (-15.8) & 14.5 \\
& 2p & n (-1.9) & n & -4.8 (-6.1) & 13.5 & -8.9 (-9.8) & 13.4 \\
& 2d & n (n) & n & n (-0.7) & n & -2.2 (-3.7) & 11.9 \\
\hline \hline
\end{tabular}}
\caption{\label{tab:phienergies} $\phi$-nucleus single-particle energies $E$
  and half widths $\Gamma/2$, for three values of $\Lambda_K$.
  When only the real part of the potential is included, the corresponding
  single-particle energy $E$ is given in parenthesis and $\Gamma=0$ for
  all nuclei. ``n'' indicates that no bound state is found. All quantities are
  given in MeV.}
\end{center}
\end{table}

\section{Summary and discussion}

We have calculated the $\phi$ meson mass and width in nuclear matter within an
effective Lagrangian approach. The in-medium kaon masses are
calculated in the QMC model, where the scalar and vector meson mean fields
couple directly to the light $u$ and $d$ quarks (antiquarks) in the $K$
($\Kbar$) mesons. At normal nuclear matter density we have found a
downward shift of the $\phi$ mass of a few percent. On the other hand, the
decay width has been broadened by an order of magnitude. We have also
calculated the $\phi$--nucleus bound state energies and absorption widths
for various nuclei by solving the Klein-Gordon equation with complex potentials.
We expect that the $\phi$ should form bound states for all four nuclei
selected, provided that the $\phi$ is produced in (nearly) recoilless
kinematics. This feature is even more obvious in the case where
the absorptive part of the potential is ignored. Given the similarity of the
binding energies and widths, the signal for the formation of the
$\phi$--nucleus bound states may be difficult to identify experimentally.
Therefore, the feasibility of observation of the $\phi$--nucleus bound states
needs further investigation, including explicit reaction cross section
estimates.


This work was partially supported by Conselho Nacional de Desenvolvimento
Cient\'{\i}fico e Tecnol\'ogico-CNPq, Grants No.~152348/2016-6 (J.J.C-M.),
No.~400826/2014-3 and No.~308088/2015-8 (K.T.), No.~305894/2009-9 (G.K.), and
No.~313800/2014-6 (A.W.T.), and Funda{\c c}\~{a}o de Amparo \`{a} Pesquisa do
Estado de S\~ao Paulo-FAPESP, Grants No.~2015/17234-0 (K.T.) and
No.~2013/01907-0 (G.K.). This research was also supported by the University of
Adelaide and by the Australian Research Council through the ARC Centre of
Excellence for Particle Physics at the Terascale (CE110001104), and through
Grant No.~DP151103101 (A.W.T.).

\end{document}